\documentclass[12pt]{article}

\usepackage[USenglish]{babel} 
\usepackage[T1]{fontenc}
\usepackage[ansinew]{inputenc}
\usepackage{verbatim}
\usepackage[sort,move]{cite}

\usepackage{lmodern} 

\usepackage{graphicx} 

\usepackage{amsmath}
\usepackage{amsthm}
\usepackage{amsfonts}

\usepackage{hyperref}

\usepackage{a4wide} 

\def\d{\delta}
\def\e{\epsilon}           
\def\g{\gamma}

\def\k{\kappa}                    
\def\l{\lambda}
\def\m{\mu}
\def\n{\nu}
\def\o{\omega}  
\def\p{\pi}                
\def\r{\rho}                                     
\def\s{\sigma}                                   
\def\t{\tau}

\def\G{\Gamma}

\def\del{\partial}              
  \let\g=\gamma \let\d=\delta \let\e=\epsilon
    \let\k=\kappa
\let\l=\lambda \let\m=\mu \let\n=\nu  \let\r=\rho
\let\s=\sigma \let\t=\tau

    \let\G=\Gamma

\def\nn{\nonumber} \def\bd{\begin{document}} \def\ed{\end{document}}
\def\ds{\documentstyle} \let\fr=\frac \let\bl=\bigl \let\br=\bigr
\let\Br=\Bigr \let\Bl=\Bigl
\let\bm=\bibitem
\let\na=\nabla
\let\pa=\partial \let\ov=\overline
\newcommand{\be}{\begin{equation}}
\newcommand{\ee}{\end{equation}}
\def\ba{\begin{array}}
\def\ea{\end{array}}
\def\ft#1#2{{\textstyle{{\scriptstyle #1}\over {\scriptstyle #2}}}}
\def\fft#1#2{{#1 \over #2}}
\def\del{\partial}
\def\sst#1{{\scriptscriptstyle #1}}
 \def\oneone{\rlap 1\mkern4mu{\rm l}}
\def\ie{{\it i.e.\ }}
\def\via{{\it via}}
\def\semi{{\ltimes}}
\def\str{{\rm str}}
\def\tr{{\rm tr}}
\def\Dm{{{D_{\sst{max}}}}}
\def\vac{ \left | 0 \right \rangle }
\def\kvac{ \left | k \right \rangle }

\def\sp{\; \; \;}

\def\bol{ \left | B (p^+) \right \rangle}
\def\bo1{ \left | B^0 (p^+) \right \rangle}

\def\bolt{ \left | B (p^+) \right \rangle_{\t}}

\def\boxl{ \left | B (x^-) \right \rangle}
\newcommand{\bea}{\begin{eqnarray}}
\newcommand{\eea}{\end{eqnarray}}

\def\<{ \langle }
\def\>{ \rangle }

\renewcommand{\floatpagefraction}{0.6}
\renewcommand{\textfraction}{0.2}

\newcommand\ca{\mathcal{A}}
\newcommand\vp{\varphi}
\newcommand\beal{\begin{align}}
\newcommand\bbone{\ensuremath{\mathbbm{1}}}

\newcommand{\eq}[1]{\begin{equation}#1\end{equation}}
\newcommand{\spl}[1]{\begin{split}#1\end{split}}
\newcommand{\al}[1]{\begin{align}#1\end{align}}
\newcommand{\subeq}[1]{\begin{subequations}#1\end{subequations}}

\newcommand{\arXividhepth}[1]{\href{http://arxiv.org/abs/#1}arXiv:{\tt #1} [hep-th]}
\newcommand{\arXividother}[2]{\href{http://arxiv.org/abs/#1}arXiv:{\tt #1} [#2]}

\newcommand{\bg}[1]{\hat{#1}}
\newcommand{\wj}{\widetilde{J}}
\newcommand{\reo}{\mathrm{Re}~\!\omega}
\newcommand{\imo}{\mathrm{Im}~\!\omega}
\newcommand{\ads}{AdS_4}
\newcommand{\mcal}{\mathcal{M}}
\newcommand{\ccal}{\mathcal{C}}
\newcommand{\ncal}{\mathcal{N}}
\newcommand{\boxedeq}[1]{
\begin{equation}
\fbox{
\rule[0.7cm]{0pt}{0pt}
$#1$
\rule[-0.45cm]{0pt}{0pt}
}
\end{equation}
}

\def\d{\text{d}}

\def\slashchar#1{\setbox0=\hbox{$#1$}           
\dimen0=\wd0                                 
\setbox1=\hbox{/} \dimen1=\wd1               
\ifdim\dimen0>\dimen1                        
\rlap{\hbox to \dimen0{\hfil/\hfil}}      
#1                                        
\else                                        
\rlap{\hbox to \dimen1{\hfil$#1$\hfil}}   
/                                         
\fi}

\def\Re           {{\rm Re\hskip0.1em}}
\def\Im           {{\rm Im\hskip0.1em}}

\newcommand{\E}{\text{\tiny E}}

\begin{document}

\begin{titlepage}

\begin{center}


\vskip 2cm
{\Large \bf New modes from higher curvature corrections in holography}


\vskip 1.25 cm {\bf Steffen Aksteiner, Yegor Korovin}
\\ {\vskip 0.5cm \it\small
Max-Planck-Institut f{\"u}r Gravitationsphysik,\\
Albert-Einstein-Institut, \\
Am M{\"u}hlenberg 1, 14476 Golm, Germany
}

{\vskip 0.2cm \small {\it E-mail: steffen.aksteiner@aei.mpg.de, jegor.korovins@aei.mpg.de} }

\end{center}

\vskip 1 cm

\begin{abstract}
\baselineskip=16pt
In gravitational theories involving higher curvature corrections the metric describes additional degrees of freedom beyond the graviton. Holographic duality maps these to operators in the dual CFT. We identify infinite families of theories for which these new modes cannot be truncated and the usual Fefferman-Graham expansion needs to be modified. New massive gravity in three dimensions and critical gravity in four dimensions are particular representatives of these families. We propose modified expansion, study the near-boundary behaviour of the metric and derive fall-off properties of the additional modes in theories involving higher derivative corrections. 
\end{abstract}

\end{titlepage}

\setcounter{tocdepth}{2}

\tableofcontents
\pagebreak









\section{Introduction}

Higher curvature corrections to general relativity are expected to play a role in quantum theory of gravity. They arise in perturbative string theory and as such may change some qualitative features of gravity familiar from general relativity. In flat space higher curvature corrections typically improve renormalisability of the theory but also make it non-unitary \cite{Stelle:1976gc, Stelle:1977ry} due to the massive modes with negative kinetic terms. 
By now the canonical example of a unitary gravitational theory is the new massive gravity (NMG) \cite{Bergshoeff:2009hq} in three dimensions. In this case the unitarity can be achieved because in three dimensions the massless graviton does not propagate local degrees of freedom. Therefore there is no harm in having kinetic term for the massless graviton with the wrong sign. With this choice of the overall sign the kinetic energy of the massive graviton is positive and unitarity is restored.

Generically one still expects that theories with (non-perturbative) higher curvature corrections have certain pathologies like ghosts, non-unitarity, acausality \cite{Camanho:2014apa, Papallo:2015rna}, etc. Nevertheless for some special theories some of these problems may not be present and then healthy physical interpretation  can be given. It is of its own interest to understand what kind of theories suffer from what kind of problems and which theories could be physically acceptable.

In this paper we study the role of higher curvature corrections in the context of the AdS/CFT correspondence. Properties of the dual CFT then can be reinterpreted on the gravity side using holographic dictionary.
More specifically we study the near-boundary expansion for the asymptotically locally AdS (AlAdS) solutions in theories involving quadratic higher curvature corrections. Analysis of the near-boundary expansion is a crucial step towards holographic renormalisation and hence understanding the dual CFT. In this paper we focus on certain subsectors of the Hilbert space in the (putative) field theory duals. 



In particular we emphasise that there are different universality classes of higher curvature corrections. One large class of them has a subsector which is identical to the Hilbert space arising in field theories dual to general relativity. But there are infinite families of theories for which the structure of the Hilbert space is qualitatively different and the results from GR cannot be directly generalised.

In the context of AdS/CFT one distinguishes between the space of classical solutions to the theory and the space of perturbations around a fixed background. The former gets mapped to the Hilbert space of the theory whereas the latter corresponds to the spectrum of excitations around a given state. In principle there is no simple relation between the asymptotics of solutions and the spectrum of the theory.

Usual considerations of unitarity rely on the study of the perturbative modes around a certain vacuum (usually pure AdS space). Small perturbations around AdS vacuum (and other states) in higher curvature gravities have been studied extensively (see e.g. \cite{Deser:2011xc} for a systematic analysis of different cases) and a variety of phenomena has been found. Generically higher order corrections lead to massive ghost modes. 
Some special theories were found, the so called critical gravities \cite{Bergshoeff:2009aq,Deser:2011xc,Lu:2011zk}, where these massive modes become massless/degenerate with the graviton. These are expected to violate unitarity and be dual to logarithmic CFTs (see \cite{Grumiller:2013at} for a review).

Also there is a significant body of work studying certain classes of solutions in theories with higher curvature corrections. Black holes have been studied for example in \cite{Garraffo:2008hu,Lu:2015cqa,Lu:2015psa}, non-relativistic backgrounds were constructed in \cite{Adams:2008zk,AyonBeato:2011qw}. In most of these papers it was realised that for certain values of parameters some special features emerge, e.g. certain functions in the metric become arbitrary. Our analysis in this paper gives a more systematic and general derivation of such special cases. 
 
Linearised approximation provides a simple way to count the number of local degrees of freedom. However there are cases when such counting is misleading. In particular linearised equations might possess some accidental symmetries which are not present in the full theory. This linearised instability occurred in several theories involving higher curvature corrections (see e.g. \cite{Maloney:2009ck}). The Hamiltonian formalism based on the analysis of constraints provides a more systematic and robust method of analysing propagating degrees of freedom. On the other hand it is technically more involved and often must be performed on the case by case basis.

One way to proceed with the Hamiltonian analysis is to eliminate local symmetries by fixing the gauge as completely as possible. This idea is at the heart of the Fefferman-Graham (or Henningson-Skenderis 
\cite{Henningson:1998gx}) type analysis for the asymptotically (locally) AdS spaces. When the gravity is described by GR this method identifies in a straightforward manner the degrees of freedom of the theory and is particularly well suited for holographic considerations. In this paper we utilise this approach to analyse the role of the higher curvature corrections for gravity in AlAdS spaces with a particular view towards AdS/CFT interpretation. One advantage of this strategy is that we are able to follow it for the most general theory of gravity involving corrections which are quadratic in the curvature.

One important aspect of any study of higher curvature corrections is the issue of the well-posed variational problem and corresponding boundary terms. Generically higher order theories propagate more degrees of freedom, meaning that it is not enough to consider simple Dirichlet problem as for ordinary GR. In the language of AdS/CFT correspondence  this just means that the metric in the bulk describes not only the stress-energy tensor on the boundary but some additional operators as well. In particular these other operators have independent sources and the variational problem in the bulk should take this into account \cite{Skenderis:2009nt,Smolic:2013gz}. One can of course avoid dealing with the variational problem if one treats higher curvature corrections perturbatively.

Let us now describe the sectors under consideration in more detail. First of all we assume asymptotic isotropy. More precisely we assume that different components of the metric diverge at the same rate near the boundary. Second, we shall switch on the source only for one of the modes, i.e. we allow for the general background metric on the boundary. In principle, since field equations are generically fourth order in derivatives, one should impose more boundary conditions. In particular one should allow for general sources for all independent modes. We will determine the fall off behaviour of these additional modes, but taking them into account in complete generality would lead us to consider many different cases. Moreover it is often consistent to switch off some of them. This is in fact the usual way to deal with the irrelevant deformations and we adopt it here. This can also be interpreted as imposing special boundary conditions. Thus we are not studying the most general asymptotic solution. But our analysis will be general enough to identify interesting cases where special care should be taken. There is also no loss of generality when the field equations are second order in derivatives, i.e. for the Lovelock family of gravities.

The paper is organised as follows. In the next section we review the AdS vacua of our model and fluctuations around them. We derive the fall off behaviour for both independent modes in the bulk and relate them to the masses of the linearised fluctuations around AdS. In the third section we proceed to the near boundary analysis of the field equations. Generically the form of the subleading terms is the same as in GR. We identify special classes of theories when the form of these terms or the form of the expansion has to be modified. We discuss and conclude in section \ref{discuss}. Some technicalities are delegated to the appendices.

\section{Vacua and linearised fluctuations around them}

Consider the action
\be
\label{action}
S = \frac{1}{16 \p G_{d+1}} \int d^{d+1}x \sqrt{-G} \Big[R + \frac{d (d-1)}{L^2} + L^2 (\l_1 R_{abcd} R^{abcd} + \l_2 R_{ab} R^{ab} + \l_3 R^2) \Big].
\ee
The particular case of Lovelock gravity is obtained for $\l_1 = \l_3 = \l_{GB},$ $\l_2 = - 4 \l_{GB}$. In four dimensions the Gauss-Bonnet (GB) term is topological. Above four dimensions the GB term is the only quadratic correction which does not produce ghosts.

The field equations can be written in the trace subtracted form
\begin{align}
\label{fieldeqs}
0 = E_{ab}= R_{ab} &+ \frac{d}{L^2} G_{ab} \\
+L^2 \Big[ &\frac{1}{d-1} \Big( -\l_1 Riem^2 - \l_2 Ric^2 - \l_3 R^2 -(2 \l_1 + \l_2 + 2 \l_3)\Box R \Big) G_{ab} \nn \\ &+ 2 \l_1 R_{acde} R_b{}^{cde} + (4 \l_1 + \l_2) \Box R_{ab} - (2 \l_1 + \l_2 + 2\l_3) D_a D_b R \nn \\ &- 2 (2 \l_1 + \l_2) R^{cd} R_{c (ab)d} - 4 \l_1 R_{ac} R_b{}^c + 2 \l_3 R R_{ab} \Big] . \nn
\end{align}
We use the radial-axial gauge for the bulk metric $G_{\m\n}$:
\be
\label{radax}
ds^2 = G_{ab} dx^{a} dx^{b}= dr^2 + \g_{ij}(r,x) dx^i dx^j.
\ee

From Einstein's equations we can derive the fall-off behaviour for the fields. Since the field equations are fourth order (except the Lovelock case) we expect that there are four independent boundary conditions one can impose on the metric. If the usual holographic interpretation is still valid then two of them should correspond to the sources in the dual field theory. Apart from the usual background metric there is a new source appearing. Naively it seems that the fall-off behaviour of the bulk metric determines the dimension of the dual operator. Since in principle it should be possible to switch both sources on we propose the ansatz
\be
\label{genansatz}
\g_{ij} = e^{2 r/l} (g_{(0)ij} + e^{-n r/l} g_{(n)ij}),
\ee   
where $l$ stands for the radius of the corresponding AdS vacuum. Without loss of generality we assume that $g_{(0)ij}$ represents the boundary metric, whereas $g_{(n)ij}$ is the source for the second operator. The fall off behaviour of the second source depends on the couplings of the theory in the bulk. At this point we do not make any assumption about the sign of $n$. We just note that if $n$ is negative than this would correspond to an irrelevant deformation and thus could be treated only infinitesimally.

Let us start by analysing the Einsteinian branch of solutions. Plugging the ansatz into the Einstein's equations and neglecting the $g_{(n)ij}$ for now we find the biquadratic equation
\be
\label{xeq}
\l L^2 x^2 - \frac{x}{4} + \frac{1}{L^2} = 0,
\ee
where
\be
x = \frac{4}{l^2}, \qquad \l = \frac{d-3}{8(d-1)} \Big(\l_1 + \frac{d}{2} \l_2 + \frac{d(d+1)}{2} \l_3\Big).
\ee
Notice that when $\l = 0$ (which is more general scenario than just pure Einstein theory) we get the solution
\be
x = \frac{4}{L^2} \implies l= L.
\ee
One recognises immediately the familiar AdS fall-off of the metric. More generally from the string theory perspective one needs all higher curvature couplings $\l_i$ to be small in order for the Planck length to be well below the AdS radius. In this case also $\l \sim 0$. We are going to treat higher curvature corrections non-perturbatively.

The algebraic equation \eqref{xeq} admits two real positive roots if and only if
\be
0 \leq \l \leq \frac{1}{64}.
\ee
The two roots are
\be
\label{xsolutions}
x_> = \frac{1 + \sqrt{1 - 64 \l}}{8 \l L^2}, \qquad x_< = \frac{1 - \sqrt{1 - 64 \l}}{8 \l L^2}
\ee
and correspond to the two possible AdS vacua of the theory.
The smaller root $x_<$ is continuously connected to the pure AdS solution of Einstein's gravity (i.e. when $\l \rightarrow 0$). It is known that in the Lovelock case only the AdS vacuum with the larger radius (smaller root $x_<$) is stable \cite{Boulware:1985wk}. 


A special case appears when the cosmological constant term is absent in the original action \eqref{action}. Then the $1/L^2$ term is absent in \eqref{xeq} and one of the vacua is necessarily flat while another one is (A)dS. Both of them are known to be stable \cite{Deser:2002jk}. If in addition the Einstein-Hilbert term is also absent and $\l=0$ then the theory admits (A)dS vacua with arbitrary curvature (see e.g. \cite{Alvarez-Gaume:2015rwa} for a particular example). This is just a consequence of underlying scale invariance.


How should one think about the two possible AdS solutions with the radii determined by \eqref{xsolutions}? A priori one could have thought that the two roots \eqref{xsolutions} describe the graviton and the second dynamical mode. We instead propose that the two roots describe two AdS vacua and that the second mode cannot be switched on unless there is some non-degenerate metric on the boundary. Thus we proceed to use the ansatz \eqref{genansatz} in order to determine the characteristic exponent for the second mode. Our proposal will be supported in the next section by the explicit computation of the subleading terms.


Now let us return to our general ansatz \eqref{genansatz} and focus on the terms of order $e^{-n r/l}$ (see \cite{Cunliff:2013en} for the special case of three-dimensional new massive gravity). From the trace equation (we use equations in the Gauss-Codazzi form as presented in Appendix \ref{GaussCodazzi}) we derive
\be
n (d+1-n)a_n\tr(g_{(n)})=0,
\ee
where the trace is taken using $g^{ij}_{(0)}$ and
\be
\label{a_n}
a_n = 1+4\frac{L^2}{l^2}\Big(-8\l +\frac{n}{2}\frac{d-n}{d-1}\m\Big),
\ee
where
\be
\label{mu}
\m = 2 \l_1 + \frac{d+1}{2}\l_2 + 2 d \l_3.
\ee
Notice that $\m$ vanishes in the Lovelock case. The $(ij)$ equations give
\be
n(n-d)\hat{a}_n g_{(n)ij} + \ldots =0,
\ee
where we omitted the terms involving $\tr(g_{(n)})$ which can be restored by comparing it to the trace equation. See also the next section for the explicit results for integer $n$. The $\hat{a}_n$ coefficient is
\be
\label{hata_n}
\hat{a}_n = 1-\frac{L^2}{l^2}\Big(32\frac{d-1}{d-3}\l + n(d-n)(4\l_1 + \l_2) + 4(2-d)\l_1 \Big).
\ee
 Finally the $(ri)$ equation results in
\be
n\hat{a}_n \nabla^j g_{(n)ij} + \ldots =0.
\ee

All these equation must be satisfied if some new independent sources can be introduced at order $n$. First of all we see that these equations have trivial solutions corresponding to $n=0$ or $n=d$. These are just the usual Einsteinian modes. However there are new solutions for $n$ when either $a_n$ or $\hat{a}_n$ vanish. These solutions indicate the dimension of the operator which is dual to the 'massive' mode. In the next section we will see that if $a_n$ or $\hat{a}_n$ vanish for small integer $n$ then the expansions of the graviton and of the massive mode mix and source each other. In such cases the usual Fefferman-Graham expansion for the graviton breaks down and should be modified by introducing logarithmic terms corresponding to explicit sources (or vacuum expectation values (VEVs)) for the other mode. In particular any analysis of correlation functions of the stress-energy tensor or Weyl anomaly based on GR-like expansions (e.g. \cite{Imbimbo:1999bj,Sen:2014nfa}) does not directly apply to these cases. It is clear that from the dual field theory perspective in these special cases the new operator has small integer dimension and thus naturally appears as a matter contribution to the Weyl anomaly or as a logarithmic partner of the stress-energy tensor. It would be interesting to modify the cohomological analysis of \cite{Imbimbo:1999bj} to incorporate these special theories.

For Lovelock gravities the $a_n$ and $\hat{a}_n$ coefficients do not depend on $n$ and there are no new modes as expected. There is however a special Lovelock theory for which the character of the Fefferman-Graham expansion drastically changes. This is the case in five (and higher) bulk dimensions when both $a_4$ and $\hat{a}_4$ vanish. We will say more about this gravitational Chern-Simons theory in the next section.


Let us for completeness review the analysis of the linearised fluctuations $h_{\m\n} =g_{\m\n} - \bar{g}_{\m\n}$ around AdS vacuum (of radius $l$). The systematic analysis of such fluctuations has been performed for instance in \cite{Deser:2011xc}. The equations of motions are
\be
\label{lineq}
c\mathcal{G}^L_{\m\n} + (2\l_1 + \l_2 + 2 \l_3)\Big(\bar{g}_{\m\n} \bar{\Box} - \bar{\nabla}_{\m} \bar{\nabla}_{\n} -\frac{d}{l^2}\bar{g}_{\m\n}\Big)R^L +(4\l_1 + \l_2)(\bar{\Box} \mathcal{G}^L_{\m\n} +\frac{d-1}{l^2}\bar{g}_{\m\n} R^L)=0,
\ee
where barred quantities are computed using the background AdS metric $\bar{g}$, we denote the Einstein tensor as
\be
\mathcal{G}_{\m\n} = R_{\m\n} - \frac{1}{2}R g_{\m\n} -\frac{d(d-1)}{2l^2}g_{\m\n}
\ee
and its linearisaion is
\be
\mathcal{G}^L_{\m\n} = R^L_{\m\n} - \frac{1}{2}R^L \bar{g}_{\m\n} +\frac{d}{l^2}h_{\m\n}.
\ee
The coefficient $c$ in front of the linearised Einstein tensor is 
\be
c = \frac{1}{L^2}\Big[1 - 2 \frac{L^2}{l^2}\Big(2(1-d)\l_1 + (d-1)\l_2 + d(d+1)\l_3 \Big) \Big].
\ee
Linearised curvatures are
\be
R^L_{\m\n} = \frac{1}{2}(\bar{\nabla^{\s}}\bar{\nabla_{\m}}h_{\n\s} + \bar{\nabla^{\s}}\bar{\nabla_{\n}}h_{\m\s} - \bar{\Box}h_{\m\n} - \bar{\nabla_{\n}}\bar{\nabla_{\m}}h), \quad R^L = - \bar{\Box}h +\bar{\nabla^{\n}}\bar{\nabla^{\m}}h_{\m\n} +\frac{d}{l^2}h.
\ee
Taking the trace of \eqref{lineq} we get
\be
\label{scalarmode}
\Big[2\m\bar{\Box} -\frac{d-1}{L^2}(1 -32\frac{L^2}{l^2}\l) \Big]R^L=0.
\ee
Notice that the mass of this mode is proportional to $a_d$ and hence directly related to the fall-off behaviour of the trace mode as discussed before. Something special happens when $\m = 0$, in particular for Lovelock gravity. Scalar mode gets eliminated from the spectrum. If in addition the second term also vanishes (which happens at the special point $\l=1/64$) then the $R^L$ is unconstrained. 
In the transverse traceless gauge the equations for perturbations simplify to
\be
\label{massivegrav}
(\l_2 + 4\l_1)\Big(\bar{\Box} + \frac{2}{l^2} -M^2 \Big) \Big( \bar{\Box} + \frac{2}{l^2}\Big)h_{\m\n} =0,
\ee
where
\begin{align}
\label{mass}
M^2 &= -\frac{1}{(4\l_1 + \l_2)L^2}\Big[1 -2\frac{L^2}{l^2}\Big(-2(d-3)\l_1 + d \l_2 + d(d+1)\l_3 \Big) \Big] \\ &= -\frac{1}{(4\l_1 + \l_2)L^2}\Big[1 -4\frac{L^2}{l^2}\Big(\frac{8(d-1)}{d-3}\l + (2-d)\l_1 \Big) \Big]=-\frac{\hat{a}_d}{(4\l_1 + \l_2)L^2}. \nn
\end{align}
\eqref{scalarmode} and \eqref{massivegrav} show that generically there are propagating massless graviton and massive spin two and spin zero modes.

Importantly, this value of the mass \eqref{mass} is exactly proportional to $\hat{a}_d$! This provides the direct link between perturbative masses around AdS and the fall off exponents of the massive modes! When the mass vanishes the new mode degenerates with the graviton and as we shall see in the next section the terms in the near boundary expansion begin to mix.

If the parameters of the theory are such that the mass \eqref{mass} vanishes one refers to the theory as critical. The black holes have vanishing entropy and mass in critical theories. Massive graviton becomes a logarithmic partner of the massless graviton. This leads to non-unitarity and the dual CFT is expected to be logarithmic.

\section{The Fefferman-Graham expansion}

In this section we solve the field equations close to the boundary. The equations in the Gauss-Codazzi form are presented in appendix \ref{GaussCodazzi}.

Below we focus on the case when the dimensions of the two operators do not differ by a small integer. This guarantees that the Fefferman-Graham expansions do not mix at leading order.

We are mostly interested in $2\leq d \leq 4$. We choose the gauge \eqref{radax} and expand the metric as inspired by GR
\be
\label{gexpansion}
g_{ij} = g_{(0)ij} + e^{-2r/l} g_{(2)ij} + e^{-3r/l} g_{(3)ij} + r e^{-4r/l} h_{(4)ij} + e^{-4r/l} g_{(4)ij} + \ldots.
\ee
$g_{(3)ij}$ is expected to appear in $d=3$ only. Note that the gauge \eqref{radax} in general is not consistent with the transverse traceless gauge which was convenient for the linearised analysis in the previous section. At this point this form of the near-boundary expansion \eqref{gexpansion} is an assumption inspired by GR. Later we shall see when this assumption breaks down. Now we proceed by analysing the field equations order by order.

The $(rr)$ equation leaves the $\tr(g_{(2)})$ undetermined due to a non-trivial cancellation. However the $\tr(g_{(2)})$ can be determined from the trace equation. The result is
\be
\label{geq}
a_2(\l_i) (l^2 R_{(0)} + 2 (d-1)\tr(g_{(2)})) = 0,
\ee
where the trace here is taken using $g_{(0)}^{ij}$ and
\be
a_2(\l_i) = 1 + 4 \frac{L^2}{l^2} \Big(\l_1 + \l_2 -\frac{d(d-5)}{2}\l_3\Big)= 1 + 4 \frac{L^2}{l^2} \Big( -8\l + \frac{d-2}{d-1}\m \Big)
\ee
is of the same form as computed in the previous section (see equation \eqref{a_n}). Note that $\m$ (defined in \eqref{mu}) vanishes for Lovelock gravities. Similarly from $(ri)$ and $(ij)$ we get
\begin{align}
&\hat{a}_2(\l_i) \Big( \nabla_{i}\tr(g_{(2)}) - \nabla^j g_{(2)ij} \Big)+ 2 \frac{L^2}{l^2}\hat{\l} \nabla_i\Big(2(d-1)\tr(g_{(2)}) + l^2 R_{(0)} \Big)=0
\end{align}
where $\nabla$ here denotes the covariant derivative with respect to $g_{(0)ij}$ and
\begin{align}
&\hat{a}_2(\l_i) \Big[ g_{(2)ij} \!-\! \frac{l^2}{d\!-\!2} \Big(\frac{R_{(0)} g_{(0)ij}}{2(d\!-\!1)} \!-\! R_{(0)ij} \Big) \Big]+ 4 \frac{L^2}{l^2}\hat{\l} \Big(2(d\!-\!1)\tr(g_{(2)}) \!+\! l^2 R_{(0)} \Big) g_{(0)ij}\!=\!0,
\end{align}
where
\be
\hat{a}_2(\l_i) = a_2(\l_i)-4d \frac{L^2}{l^2} \hat{\l} \qquad \text{and} \qquad \hat{\l} = \l_1 + \l_2 + 3 \l_3.
\ee
Notice that $\hat{\l}$ vanishes identically for Lovelock gravities.

Now we immediately see important differences with respect to general relativity. There are several cases to consider.

\begin{itemize}

\item

\textbf{In the case $a_2(\l_i) \neq 0$ and $\hat{a}_2(\l_i)\neq 0$} the solution to the equations above coincides with the well known result for GR:
\begin{align}
\label{trg2}
\tr(g_{(2)}) &= -\frac{l^2 R_{(0)}}{2(d-1)}, \\
\label{divg2}
 \nabla^j g_{(2)ij} &= \nabla_{i}\tr(g_{(2)}), \\
 \label{g2}
g_{(2)ij} &= \frac{l^2}{d-2} \Big(\frac{R_{(0)}}{2(d-1)}g_{(0)ij} - R_{(0)ij} \Big),
\end{align}

where the last equation holds only if $d \neq 2$. It is clear however that for certain combinations of the higher curvature couplings $\l_i$ (some of) the equations above leave some components of $g_{(2)ij}$ undetermined!

\item
\textbf{In the case $a_2(\l_i) \neq 0$ and $\hat{a}_2(\l_i)= 0$} the trace $\tr(g_{(2)})$ is still given by \eqref{trg2} however $\nabla^j g_{(2)ij}$ and $g_{(2)ij}$ are left undetermined by the near boundary analysis. This is in contrast to GR where these components get expressed algebraically in terms of boundary data as in \eqref{divg2} and \eqref{g2}. It is clear that this arbitrariness is due to the appearance of the new mode at this order as discussed in the previous section: the condition $a_2(\l_i) \neq 0$ makes it traceless and that is why the trace of the $g_{(2)}$ does not get modified. However the traceless part of the new mode is arbitrary due to $\hat{a}_2(\l_i)= 0$. Presumably one would need to introduce a logarithmic mode at this order in order to incorporate this new mode. This case is inconsistent with Lovelock condition for which $\hat{a}_2(\l_i) = a_2(\l_i)$.

\item
\textbf{In the case $a_2(\l_i) = 0$ and $\hat{a}_2(\l_i)\neq 0$} the trace $\tr(g_{(2)})$ is left undetermined, whereas $\nabla^j g_{(2)ij}$ and $g_{(2)ij}$ are expressed in terms of $\tr(g_{(2)})$:
\begin{align}
\nabla^j g_{(2)ij} &= \frac{1}{d}\nabla_{i}\Big(\tr(g_{(2)}) -\frac{l^2}{2}R_{(0)}\Big), \\
g_{(2)ij} &= \frac{1}{d(d-2)} \Big((d-2)\tr(g_{(2)})g_{(0)ij} +l^2 (R_{(0)}g_{(0)ij} - d R_{(0)ij}) \Big).
\end{align}
Again, this case cannot appear for Lovelock gravities.

Interestingly the critical point of the so-called new massive gravity (NMG) in three bulk dimensions \cite{Bergshoeff:2009hq} is a particular member of this special family with
\be
\l_1=0, \qquad \l_2 = 1, \qquad \l_3 = -\frac{3}{8}, \quad \text{so that} \quad \l = 1/64.
\ee
It is also known that in NMG there is enhanced gauge symmetry on the level of linearised field equations around (A)dS vacuum \cite{Bergshoeff:2009aq}. This gauge invariance is of an unusual type and gives rise to partially massless (PM) fields \cite{Deser:1983mm, Deser:2001us}. More concretely the gauge parameter in this case is a scalar and removes one degree of freedom - the trace of the metric. At the same time the equation for fluctuations degenerates and logarithmic modes appear. In this case the 'partially massless gravity' is conjectured to be dual to logarithmic CFT \cite{Grumiller:2010tj}. However this Weyl invariance cannot be promoted to interacting theory \cite{Blagojevic:2011qc,Hohm:2012vh}. Moreover there is a no-go theorem prohibiting having interacting PM fields coupled to gravity in a unitary theory \cite{Dolan:2001ih, Joung:2014aba} and it is true that NMG at the critical point is non-unitary. Our analysis confirms that there is no non-linear extension of this linearised Weyl symmetry and what appears as a gauge freedom is just the presence of an additional boundary condition for the trace mode. 


\item
\textbf{Finally in the case $a_2(\l_i) = 0$ and $\hat{a}_2(\l_i)= 0$} all the components of $g_{(2)ij}$ are left undetermined. For generic number of boundary dimensions $d$ these conditions define a one-parameter family of theories (the independent parameter can be conveniently chosen to be $\l_3$). Notice however that this last case cannot be realised if $d=2$ or $d=3$.

\end{itemize}

For $d=2$ everything gets determined by a single parameter $\hat{\l}$. As a result either $\tr(g_{(2)})$ or $\nabla^j g_{(2)ij}$ (or both) is (are) determined and $g_{(2)ij}$ is left arbitrary by the near-boundary analysis as expected. Lovelock term does not modify the result with respect to GR.




Let us now move on to the next order in the Fefferman-Graham expansion. For $g_{(3)ij}$ we obtain:
\begin{align}
0 &= a_{3} \tr(g_{(3)}), \\
0 &= \hat{a}_{3} (\nabla_i \tr(g_{(3)}) - \nabla^j g_{(3)ij}) + 2 \frac{L^2}{l^2} \k \nabla_i \tr(g_{(3)}),\\
0 &= \hat{a}_{3}((d-3)g_{(3)ij} + \tr(g_{(3)}) g_{(0)ij}) + \frac{2 (2d-3)}{d-1} \frac{L^2}{l^2} \k \tr(g_{(3)}) g_{(0)ij},
\end{align}
where
\begin{align}
&a_{3} = 1 + \frac{d-3}{d-1}\frac{L^2}{l^2}\Big(8 \l_1 + (d+3)\l_2 - 2d(d-5)\l_3\Big) = 1 + 4\frac{L^2}{l^2}\Big(-8\l +\frac{3}{2}\frac{d-3}{d-1}\m \Big), \\ &\hat{a}_3 = a_3 - \frac{2d}{d-1}\frac{L^2}{l^2} \k, \quad\text{with} \quad \k = 4(d-3)\l_1 + (3d-7)\l_2 + 8(d-2)\l_3,
\end{align}
once again in agreement with out general expressions \eqref{a_n} and \eqref{hata_n}. Gor general $d$ the analysis of different cases is analogous to that at the previous order. There is an important difference however in $d=3$. Recall that in $d=3$ the parameter $\l$ vanishes and there is unique AdS vacuum with radius $l=L$. Moreover $a_3$ is identically equal to one and hence $\tr(g_{(3)})$ is forced to vanish, confirming the fact that in $d=3$ there is no Weyl anomaly. Nevertheless it is still possible for $\hat{a}_3$ to vanish. In any case $\l_1$ drops out of the analysis due to the ($d-3$) prefactor. This presumably corresponds to the fact that in $AdS_4$ the non-dynamical Euler density can be added to the action to remove the $\l_1 Riem^2$ term. 

Actually there exists a well-known example of a theory for which $\hat{a}_3$ vanishes. This is the logarithmic point of the so called critical gravity in four bulk dimensions \cite{Lu:2011zk} for which
\be
\l_1=0, \quad \l_2 = -3\l_3 = \frac{3}{d(d-1)}.
\ee
In this case it is known that the graviton acquires a logarithmic partner and the near-boundary expansion has to be modified by logarithmic terms.

Now we move to the next order. In $AdS_5$ one has a logarithmic term $h_{(4)ij}$ already in GR. With the higher curvature corrections we get for $h_{(4)ij}$ the following equations:
\begin{align}
&(d-3) a_4(\l_i) \tr(h_{(4)}) = 0, \\
&\hat{a}_4(\l_i) (\nabla_i \tr(h_{(4)}) - \nabla^j h_{(4)ij}) + 4 \frac{L^2}{l^2}\n \nabla_i \tr(h_{(4)}) =0, \\
&\hat{a}_4(\l_i)\Big[(d-4)h_{(4)ij} +\tr(h_{(4)})g_{(0)ij}  \Big] + 8\frac{d-2}{d-1} \frac{L^2}{l^2}\n \tr(h_{(4)})g_{(0)ij} =0,
\end{align}
where
\begin{align}
a_4(\l_i) &= 1 + \frac{2}{d-1} \frac{L^2}{l^2} \Big(2 (3d -13) \l_1 + (d^2 - 3 d - 8) \l_2 - d (d^2 - 10 d + 29) \l_3 \Big) \\ &=
 1 + 4 \frac{L^2}{l^2} \Big( -8\l + 2\frac{d-4}{d-1}\m \Big) \nn
\end{align}
and
\be
\hat{a}_4(\l_i)=a_4(\l_i) - 4 \frac{d}{d-1} \frac{L^2}{l^2}\n\quad \text{with} \quad \n = (3d-13)\l_1 + (2d-7)\l_2 + 5 (d-3)\l_3.
\ee

Before we present the results for $g_{(4)ij}$ let us introduce some notation by reminding the results for GR. The equations one obtains in GR are
\begin{align}
0&=\mathcal{E}= 4(d-3)\tr(g_{(4)}) + (5-2d)\tr(g^2_{(2)}) - \tr(g_{(2)})^2  - l^2 R_{(0)}^{ij}g_{(2)ij} \\&+ (7-d) \tr(l h_{(4)})- l^2 \gamma, \nn \\
0&=\mathcal{E}_i=2(\nabla_i \tr(g_{(4)}) - \nabla^j g_{(4)ij}) - \frac{3}{4}\nabla_i  \tr(g^2_{(2)}) - \frac{1}{2}g_{(2)ij} \nabla^j\tr(g_{(2)}) + \nabla^j (g^2_{(2)})_{ij}\\
& - \frac{1}{2}(\nabla_i \tr(h_{(4)}) - \nabla^j h_{(4)ij}), \nn \\
0&=\mathcal{E}_{ij} ={2(d-4)} g_{(4)ij} + {2} (g_{(2)}^2)_{ij} +\tr(2 g_{(4)} - g^2_{(2)})g_{(0)ij}+ \frac{(8-d)}{2} lh_{(4)ij}\\ &+l^2\Big(\nabla^k \nabla_{(i}g_{(2)j)k} - \frac{1}{2}\nabla_i \nabla_j \tr(g_{(2)}) - \frac{1}{2} \Box g_{(2)ij}\Big), \nn
\end{align}
where
\be
\g = \Box \tr(g_{(2)}) - \nabla^i \nabla^j g_{(2)ij}.
\ee

Now in the presence of higher curvature corrections \textbf{and if the result for $g_{(2)ij}$ is the same as in GR} (i.e. if $a_2 \neq 0$ and $\hat{a}_2 \neq 0$ ) we get
\begin{align}
\label{trg4}
0
&= a_4(\l_i) \mathcal{E}  + \frac{d-3}{d-1}l^2 L^2 \l_1 Weyl_{(0)}^2 -8\frac{L^2}{l^2}\frac{(d-8)(d-3)}{d-1} \m \tr(l h_{(4)}),  \\
0 &= \hat{a}_4(\l_i) \mathcal{E}_{i} + 2\frac{L^2}{l^2} \n \nabla_i\tr(4 g_{(4)} - g_{(2)}^2)-2\frac{L^2}{l^2}(d-8)(4\l_1 + \l_2)\nabla^j h_{(4)ij} \\ &+2\frac{L^2}{l^2} \Big((3d-13)\l_1 - 3(d-5)\l_2 + (47-9d)\l_3 \Big) \nabla_i \tr(l h_{(4)}) \nn \\
&+\frac{2L^2 l^2}{(d-2)^2} \l_1 \Big[\nabla_i \Big(Ric^2 -\frac{1}{2(d-1)}R^2\Big) + R_{ij}\nabla^j R -2R^{jk}\nabla_k R_{ij} + 2(d-2)R_{ijkl}\nabla^l R^{jk} \Big], \nn\\
 0 &= \hat{a}_4(\l_i) \mathcal{E}_{ij} + 2{L^2}{l^2}  \l_1 \Big((Weyl^2)_{ij} - \frac{1}{2(d-1)}(Weyl^2)g_{(0)ij} \Big) \\&
+\frac{4(d-2)}{d-1} \frac{L^2}{l^2}\n \tr(4 g_{(4)} - g_{(2)}^2) g_{(0)ij}+2\frac{L^2}{l^2}(d-8)(d-4)(4\l_1 + \l_2)h_{(4)ij} \nn \\
&+\frac{L^2}{l^2}\frac{2}{d\!-\!1}\Big[2(5d^2\!-\!47d+100)\l_1 \!+\!(7d^2\!-\!61d\!+\!112)\l_2 \!+\! 2(9d^2\!-\!75d \!+\! 124)\l_3 \Big]\tr(l h_{(4)})g_{(0)ij}.\nn
\end{align}

The discussion here is similar to that we had at the previous orders. Again there are four cases to consider depending on whether $a_4(\l_i)$ and/or $\hat{a}_4(\l_i)$ are zero or not.
\begin{itemize}
\item
$a_4(\l_i) \neq 0$, $\hat{a}_4(\l_i) \neq 0$: the result is the same as in GR. 
\item
$a_4(\l_i) = 0$, $\hat{a}_4(\l_i) \neq 0$: the trace of $g_{(4)}$ is not determined while the $\nabla^j g_{(4)ij}$ and the transverse traceless part are determined in terms of the trace in $d>4$ (in $d=4$ instead of $g_{(4)ij}$ the $h_{(4)ij}$ is determined). 
\item
$a_4(\l_i) \neq 0$, $\hat{a}_4(\l_i) = 0$: the trace is determined while the divergence and transverse traceless part are arbitrary.
\item
$a_4(\l_i) = 0$, $\hat{a}_4(\l_i) = 0$: nothing gets determined.
\end{itemize}

In a generic theory the only non-trivial correction to the trace of $g_{(4)ij}$ is proportional to $\l_1 Weyl_{(0)}^2$. For four dimensional dual CFT this trace should give the trace (or Weyl) anomaly (the exact expression however depends on the counterterm action). Thus we find a shift in the $c$ central charge of the trace anomaly. Interestingly only the $Riem^2$ term in the action contributes to this shift. This is consistent with the older result by \cite{Blau:1999vz, Schwimmer:2003eq}. Actually it is easy to see that only $\l_1$ can destroy the equality between the $c$ and $a$ anomaly coefficients. This is due to the fact that the usual Weyl anomaly $E_4 - Weyl^2$ (here $E_4$ is the Euler density in four dimensions) involves the Ricci tensor only. Therefore $c$ and $a$ can be non-equal to each other only if the anomaly gets modified by a square of the Riemann tensor. This explains the special role of the $\l_1$ coefficient. Curiously the $Riem^2$ term is also the only one which cannot be brought back to GR (plus some matter) by a local field redefinition.

However, when $a_4(\l_i) = 0$, the $\tr(g_{(4)})$ is not determined and $\tr(h_{(4)})$ is determined instead. Something interesting happens when $\m = a_4(\l_i) = 0$. In this case the equation \eqref{trg4} appear to be inconsistent unless $\l_1 Weyl_{(0)}^2 =0$. However when $\m = a_4(\l_i) = 0$ the form of the expansion should have been modified already at the order of $g_{(2)ij}$ by allowing new logarithmic terms. It seems that the two expansions begin to mix and the background metric (or more precisely the background Weyl tensor) constrains the new mode.





To make contact with something familiar, notice that in Lovelock gravities the $\m$ coefficient vanishes and $a_4(\l_i) =a_3(\l_i)= a_2(\l_i)$. If moreover $\l$ is such that $a_4(\l_i)=0$ (which in $5$ bulk dimensions corresponds exactly to the Chern-Simons gravity) than the coefficients $g_{(2)ij}$, $h_{(4)ij}$ and $g_{(4)ij}$ are not determined by the field equations! This degeneracy for Chern-Simons gravity appeared in the literature before \cite{Wheeler:1985qd, Charmousis:2002rc}. Moreover this arbitrariness is not due to some unidentified boundary condition since the field equations are of second order. In fact there exist (asymptotically AdS) solutions involving unconstrained functions of coordinates, i.e. field equations do not fix the metric entirely. Notice however that such solutions necessarily have vanishing mass. Also the effective action for the fluctuations around such solutions do not have standard quadratic terms. It is conceivable that this degeneracy is related to the gauge symmetry enhancement, however the rigorous (canonical) count of the number of degrees of freedom in the degenerate case cannot be performed by the standard methods \cite{Banados:1995mq, Banados:1996yj,Miskovic:2005di}.

The Chern-Simons gravity belongs to the Lovelock family of gravities, i.e. the field equation are second order in derivatives. Generically the field equations set some products of curvature two-forms to zero:
\be
\epsilon_{a_1 \ldots a_{d+1}}(R^{a_1 a_2} + l^{-1}e^{a_1} e^{a_2}) \ldots (R^{a_{d-1} a_d} + l^{-1}e^{a_{d-1}} e^{a_d}) = 0,
\ee
where $a_i$ stands for tangent space index. When
\be
R^{a_1 a_2} + l^{-1}e^{a_1}e^{a_2} = 0
\ee
is satisfied the quadratic part of the action for the fluctuations is vanishing, and thus there is no propagation around such background in perturbative sense. In particular this results in the families of solutions involving arbitrary functions. In Chern-Simons theory based on gauge connection such solutions correspond to the unbroken phase, i.e. they are gauge equivalent to the symmetric (non-geometric) background with $e = \o = 0$. However such gauge transformations make the vielbein $e$ non-invertible and hence are forbidden in gravitational theory.

Technically speaking, some of the Hamiltonian constraints become dependent on such degenerate backgrounds. In fact it is known \cite{Banados:1995mq, Banados:1996yj} that the constraint coming from radial reparametrization invariance\footnote{In \cite{Banados:1995mq, Banados:1996yj} the backgrounds of the form $\mathbb{R} \times M_4$ were considered and $\mathbb{R}$ was referred to as 'time' direction. In the present context $\mathbb{R}$ corresponds to the radial direction.} is not independent from other constraints. Actually the AdS solution is a maximally degenerate background, i.e. the symplectic form vanishes at this point in phase space and there are no local degrees of freedom propagating around such backgrounds. Importantly, expanding the theory around pure (A)dS vacuum we find that there is no quadratic piece in the action and thus the concept of the propagation is not well defined (in the perturbative sense) \cite{Chamseddine:1990gk}. We encountered related phenomenon when we observed that near boundary analysis leaves coefficients in the expansion arbitrary. It would be of great theoretical interest to understand better the canonical structure of the gravitational Chern-Simons theory in $5d$ and the interpretation of this exotic gravitational theory in the AdS/CFT context.

\section{Conclusions}
\label{discuss}

In this paper we have studied the influence of the higher curvature corrections on the form of the near-boundary expansion of the metric in asymptotically locally AdS spaces. Our starting point of this analysis is the ansatz
\be
\label{ans}
\g_{ij} = e^{2 r /l}(g_{(0)ij} + e^{-n r /l} g_{(n)ij} ),
\ee
where $g_{(0)ij}$ is the background metric, whereas $g_{(n)ij}$ is the source (or the VEV) of the new operator. The theory determines for us the possible values of $l$ and $n$. The characteristic exponent $n$ is linked to the mass of massive mode around the AdS vacuum. We emphasise that both these parameters have to be determined simultaneously since one necessarily needs a non-degenerate metric $g_{(0)ij}$ in order to introduce physical position-dependent couplings $g_{(n)ij}$. This ansatz serves as a seed for subsequent determination of subleading terms. Clearly this analysis can become intricate due to the mixing of the Einsteinian and the new mode.

For a general theory involving corrections which are quadratic in the Riemann tensor we identified the fall-off behaviour of the additional modes. There are infinite classes of theories for which the new modes mix with the GR-like mode. In our explicit analysis this phenomenon manifests itself as arbitrariness of certain subleading terms. Relatively well understood examples of the special cases are the new massive gravity in 3d and critical gravity in 4d. Partial holographic dictionary in these cases has been established in \cite{Alishahiha:2010bw, Johansson:2012fs}.

The theories with arbitrary terms in the near boundary expansion provide counterexample to a statement that this near boundary terms are completely universal \cite{Imbimbo:1999bj}. They are universal only if there are no operators with small integer dimension which spoil the cohomological analysis. The methods of computing correlations functions or Weyl anomaly assuming this universal behaviour do not apply directly to the special cases.

In the $AdS_5$ case we found that if the logarithmic modes do not appear then only $\l_1$ coefficient shifts the $c$ central charge. In particular we found that the '$R^2$' anomaly - characteristic feature of scale but not conformally invariant theories - does not appear in the trace anomaly in QFTs dual to gravity with quadratic curvature corrections.  This extends the old result of \cite{Henningson:1998gx} beyond pure GR.

This identification of the space of asymptotic solutions is just a first step in the program of holographic renormalisation. The next obstacle on the way is the well-posedness of the variational problem. Our results should be helpful in this direction. In the ansatz \eqref{ans} the sources of the boundary theory are manifest and hence exactly these terms in the near-boundary expansion should be held fixed in the variational problem. The bulk-covariant form of the last statement would allow us to determine necessary boundary terms. For now this problem remains open.



For Lovelock gravities arbitrary coefficients appear only at one special point: the gravitational Chern-Simons theory. In the Chern-Simons case the ambiguity in the expansion coefficients can be partly understood from the degeneracy of the AdS vacuum. For generic Lovelock theory there are two maximally symmetric vacua with different radii. The two radii coincide at the Chern-Simons point. Viewing Fefferman-Graham expansion as a perturbation around AdS vacuum we see that the degeneracy at the Chern-Simons point is due to the fact that there are two branches of solutions emanating from the doubly degenerate vacuum. Clearly starting from this degenerate solution there is no unique way to extend it into the bulk. It would be of great theoretical interest to see how exotic properties of five-dimensional Chern-Simons theory reflect themselves in the dual field theory.



\section*{Acknowledgments}
We would like to thank Kostas Skenderis for suggesting the project and collaboration at initial stages. We are also grateful to Xian Camanho and Stefan Theisen for useful discussions. The majority of computations in this paper have been verified using the xAct package \cite{xAct} for Mathematica.

\appendix

\section{Some technical details}

Here we collect some technical results.

In the formulas below the Christoffel symbols are associated with the bulk metric $G_{ab}$, indices from the beginning of the alphabet $a,b,c,e,f$ refer to the bulk coordinates $(r,x_i)$, while the indices from the middle of the alphabet ${i,j}$ refer to the boundary directions only.

In the gauge \eqref{radax} the extrinsic curvature is given by
\be
K_{ij} = \frac{1}{2}\g'_{ij},
\ee
where prime denotes the radial derivative. The Christoffel symbols are
\begin{align}
\G^r_{rr} = \G^r_{ri} = \G^i_{rr} =0; \qquad \G^r_{ij} = -\frac{1}{2}\g'_{ij} = - K_{ij}; \qquad \G^i_{rj} = K^i{}_j;\qquad \G^{i}_{jk}[G] =  \G^{i}_{jk}[\g].
\end{align}
We note the following useful relation:
\be
D_r K_{ij} = \pa_r K_{ij} - 2 (K^2)_{ij}
\ee
Gauss-Codazzi decomposition of the Riemann tensor is 
\begin{align}
R_{ijkl}[G] &= R_{ijkl}[\g] + K_{jk}K_{il} - K_{ik} K_{jl}, \\
R_{rijk}[G] &= \nabla_k K_{ij} - \nabla_j K_{ik}, \\
R_{rirj}[G] &=-D_r K_{ij} - K_{ik} K^k{}_j= -K'_{ij} + (K^2)_{ij},
\end{align}
where $\nabla$ here denotes the covariant derivative w.r.t. $\g$.
Similarly the components of the Ricci tensor are
\begin{align}
R_{ij}[G] &= R_{ij}[\g] - K'_{ij} + 2 (K^2)_{ij} - \tr(K) K_{ij}, \\
R_{ri}[G] &= \nabla^j K_{ij} - \nabla_i \tr(K), \\
R_{rr}[G] &= \tr( K^2 - K') = - \pa_r\tr(K) - \tr(K^2),
\end{align}
where the trace here is taken with $\g^{ij}$. Finally,
\begin{align}
R[G] = R[\g] +\tr(3 K^2 - 2 K')- \tr(K)^2 = R[\g] - \tr(K^2) - \tr(K)^2 - 2 \pa_r\tr(K),
\end{align}
where we have used
\be
\tr(K') = \pa_r\tr(K) + 2 \tr(K^2).
\ee
We shall also need
\begin{align}
Ric^2[G] &= Ric^2[\g] + 2 \nabla_n K^{in}\Big(\nabla^j K_{ij} - 2 \nabla_i \tr(K)\Big) + 2 \nabla_i \tr(K) \nabla^i \tr(K) \\ &+ 2 R^{ij}[\g]\Big( - K'_{ij} + 2(K^2)_{ij} - \tr(K) K_{ij} \Big) + \tr(K'^2 - 4 K' K^2 + 4 K^4) \nn \\ &+ \tr(K') \tr(K'- 2 K^2) + 2 \tr(K) \tr(K K' - 2 K^3) + \tr(K^2) \Big( \tr(K^2) + \tr(K)^2\Big), \nn
\end{align}
and
\begin{align}
Riem^2[G] &= Riem^2[\g] +2 R^{ijkl}[\g](K_{jk} K_{il} - K_{ik} K_{jl}) \\ &+ 8 \nabla^k K^{ij} \Big( \nabla_k K_{ij} -  \nabla_j K_{ik}\Big)  + 2 \tr(K^2)^2 + 2 \tr(K^4 + 2 K'^2 - 4 K' K^2), \nn
\end{align}
where we use the notation
\be
\tr(K'^2) = K'_{ij} K'_{lm} \g^{il} \g^{jm}
\ee
and similarly for other traces. Also
\begin{align}
\Box_G R[G] &= \Big( \pa_r^2 + \tr(K) \pa_r + \Box_{\g} \Big) \Big[ \tr(3 K^2 - 2 K') - \tr(K)^2 + R[\g] \Big]. \nn
\end{align}
Next we move to analyse the terms $R_{acde} R_b{}^{cde}$:
\begin{align}
R_{icde}[G] R_{j}{}^{cde}[G] &\!=\! \frac{1}{2} R_{iklm}[\g] R_{j}{}^{klm}[\g] + 2 K^{mn}K_j{}^l R_{imnl}[\g] + \nabla_i K^{mn} \nabla_j K_{mn} \\
&+ \nabla^n K_i{}^m \Big(2 \nabla_n K_{jm} - \nabla_m K_{jn} - 2 \nabla_j K_{mn} \Big)  + \tr (K^2) (K^2)_{ij} \nn \\
&-2  (K^2 K')_{ij}  + \g^{mn} K'_{in} K'_{jm} + (i \leftrightarrow j), \nn \\
R_{icde}[G] R_{r}{}^{cde}[G] & =  2 \Big[ R_{iklm}[\g] \nabla^m K^{kl} + K_i{}^l K^{mn}(\nabla_k K_{mn} - \nabla_n K_{lm}) \\ & \qquad + (K' -K^2)_{mn} (\nabla_i K^{mn} - \nabla^m K_i{}^n) \Big]. \nn \\
R_{rcde}[G] R_{r}{}^{cde}[G] & = 2 \Big[ \nabla^l K^{ij} (\nabla_l K_{ij} - \nabla_j K_{il}) + \tr(K'^2 - 2 K^2 K' + K^4) \Big].
\end{align}
The next term we analyse is $\Box_G R_{ab}[G]$. The general formula is (from now on we drop the argument of the Riemann curvatures - the reader can easily figure out which metric is meant from the form of indices)
\begin{align}
\Box_G R_{ab} = \frac{1}{2} D_r D_r R_{ab} + \g^{ij} \Big[& \frac{1}{2}\pa_i \pa_j R_{ab} -\frac{1}{2} \G_{ij}^e \pa_e R_{ab}- R_{be} \pa_j \G^e_{ai} - 2 \G_{ai}^e \pa_j R_{be} \\ & + \G^e_{ij} \G^f_{ae} R_{bf} + \G^e_{aj}(\G_{ie}^f R_{bf} + \G^f_{ib} R_{ef}) \nn\Big] + (a \leftrightarrow b),
\end{align}
where
\be
D_r D_r R_{ab} = \frac{1}{2}\pa_r^2 R_{ab} - R_{bc} \pa_r \G^c_{a r} - 2 \G^e_{ar} \pa_r R_{be} + \G^e_{ar} (\G^c_{r e} R_{bc} + \G^c_{rb} R_{ce}) + (a \leftrightarrow b).
\ee
The particular components are 
\begin{align}
\Box_G R_{rr} &\!=\! \Big( \pa_r^2 + \tr(K) \pa_r + \Box_{\g} -
 2 \tr(K^2) \Big)R_{rr} \!-\! (2 \nabla_i K^{ij}  + 4 K^{ij} \nabla_i) R_{r j} + 2 (K^2)^{ij} R_{ij}, \\
\Box_G R_{ri} &\!=\! \Big( D_r D_r \!+\! \tr(K) D_r + \Box_{\g} \!-\! \tr(K^2) \Big)R_{ri} + (\nabla^j K_{ji} + 2 K_{ij} \nabla^j)R_{rr} \!-\! 3 (K^2)_i{}^j R_{r j} \\ 
&- (\nabla_j K^{jl} + 2 K^{jl} \nabla_j) R_{il}, \nn \\
\Box_G R_{ij} & = \frac{1}{2} (D_r D_r + \tr(K) D_r + \Box_{\g})R_{ij} \\
&+ (K^2)_{ij} R_{rr} + (\nabla^n K_{jn} + 2 K_j{}^n \nabla_n)R_{ri}  -(K^2)_j{}^n R_{in} +  (i \leftrightarrow j). \nn
\end{align}
It is useful to note that
\begin{align}
D_r R_{ij} &= \frac{1}{2} \pa_r R_{ij} - K_i{}^n R_{jn} + (i \leftrightarrow j), \\
D_r D_r R_{ij} &= \frac{1}{2} \pa^2_r R_{ij}  + K_i{}^n K_j{}^m R_{mn} + \Big((K^2)_i{}^n - \pa_r K_i{}^n - 2 K_i{}^n \pa_r \Big) R_{jn} + (i \leftrightarrow j).
\end{align}
In these formulas we implicitly assume that the $(d+1)-$dimensional curvatures ($R[G]$) has been decomposed into $d-$dimensional ones ($R[\g]$).

The terms $D_a D_b R$:
\begin{align}
D_r D_r R &= \pa_r \pa_r R, \\
D_i D_r R &= \nabla_i \pa_r R - K_i{}^j \nabla_j R, \\
D_j D_i R & = \nabla_j \nabla_i R + K_{ij} \pa_r R.
\end{align}
The next term is $R_{a c} R_b{}^c$:
\begin{align}
R_{r c} R_r{}^c &= \Big(\tr(K' - K^2) \Big)^2 + \nabla_i \tr(K) \nabla^i \tr(K) + \nabla_j K^{ij} (\nabla^n K_{in} - 2 \nabla_i \tr(K)), \\
R_{i c} R_r{}^c &= \Big(\tr(K' - K^2) \Big)  (\nabla_i \tr(K) - \nabla^j K_{ij}) \\&+ (R_{ij} - K'_{ij} + 2(K^2)_{ij} - \tr(K) K_{ij}) (\nabla_n K^{jn} - \nabla^j \tr(K)), \nn \\
R_{ic} R_{j}{}^c & = \frac{1}{2} \Big (\nabla^m K_{im} \nabla^n K_{jn} + \nabla_i \tr(K) \nabla_j \tr(K) + (R^2)_{ij} + \g^{mn} K'_{im} K'_{jn} \Big) +2(K^4)_{ij} \\ & -2\tr(K) (K^3)_{ij} + \frac{1}{2} \tr(K)^2 (K^2)_{ij} -\nabla_i \tr(K) \nabla^n K_{jn} - (R K')_{ij} + 2 (R K^2)_{ij} \nn \\&- \tr(K) (R K)_{ij} - 2 (K' K^2)_{ij} + \tr(K) (K K')_{ij} +  (i \leftrightarrow j). \nn
\end{align}
Finally, the $R^{cd} R_{c(ab)d}$ terms give:
\begin{align}
R^{cd} R_{crrd} &= \tr(R K' - R K^2 + 3 K^2 K' - 2 K^4 - K' K') + \tr(K) \tr(K^3 - K K'), \\
R^{cd} R_{c(ri)d} & = (\nabla^j \tr(K) - \nabla_n K^{jn}) (K'_{ij} - (K^2)_{ij}) \\ & \quad + (R_{mn} -K'_{mn} + 2(K^2)_{mn}- \tr(K) K_{mn}) (\nabla_i K^{mn} - \nabla^n K_i{}^m), \nn \\
R^{cd} R_{c(ij)d} & = \tr( K^2 \!-\! K') (K'_{ij} \!-\! (K^2)_{ij}) +(K_i{}^m K_j{}^n + R_i{}^m_j{}^n)(K'_{mn} - 2 (K^2)_{mn} - R_{mn}) \\
& \quad+ \tr(K)(K^{mn} R_{minj} +(K^3)_{ij}) - 2(\nabla^n K_{ij} - \nabla_{(i} K_{j)}{}^n) (\nabla_n \tr(K) - \nabla^m K_{mn}) \nn  \\ & \quad+ \tr(2 K^3 - K K' + KR) K_{ij} - \tr(K) \tr(K^2) K_{ij}. \nn
\end{align}
We expand the metric (in $d=4$) as
\begin{align}
\g_{ij} &= e^{2 r /l} \Big[g_{(0)ij} + e^{-2r/l} g_{(2)ij} + r e^{-4r/l} h_{(4)ij} + e^{-4r/l} g_{(4)ij} + \ldots \Big], \\
\g^{ij} &=  e^{-2 r /l} \Big[g_{(0)}^{ij} - e^{-2r/l} g_{(2)}^{ij} - r e^{-4r/l} h_{(4)}^{ij} + e^{-4r/l} (g_{(2)} g_{(2)} -g_{(4)})^{ij} + \ldots \Big],
\end{align}
where the indices of $g_{(a)}$ and $h_{(a)}$ are raised and lowered using $g_{(0)}$.
The extrinsic curvature then is
\begin{align}
K_{ij} &= \frac{1}{l} e^{2 r /l} \Big[g_{(0)ij} - r e^{-4r/l} h_{(4)ij} + e^{-4r/l} (\frac{l}{2} h_{(4)ij} - g_{(4)ij}) + \ldots \Big].
\end{align}

\section{Gauss-Codazzi decomposition of the field equations}
\label{GaussCodazzi}

The $(rr)$ component of the Einstein equations \eqref{fieldeqs} now reads
\begin{align}
&0= \tr(K^2 - K') + \frac{d}{L^2} \\
& +L^2 \Big[-\frac{\l_1 Riem_{\g}^2 + \l_2 Ric_{\g}^2}{d-1} + \Big(\frac{2 (2-d) \l_2}{d-1} - 4 \l_1\Big) \tr(R K') + \Big(12 \l_1 + \frac{4 (d-2) \l_2}{d-1} \Big) \tr(R K^2)  \nn \\
& +\frac{2 \l_2}{d-1} \tr(K) \tr(R K) - \frac{4 \l_1}{d-1} R^{ijmn} K_{jm} K_{in} -\frac{\l_3}{d-1} \Big(R^2 + 2 R(\tr(3 K^2 - 2 K') - \tr(K)^2) \Big) \nn \\ & + \Big(\frac{4(d-2) \l_3 -\l_2}{d-1} - 4\l_1 \Big) \tr(K')^2 + \frac{(28d-30)\l_1 + 4(2d-3)\l_2}{d-1} \tr(K^4) -\frac{\l_3}{d-1} \tr(K)^4 \nn \\
& +\frac{(8d-12)\l_1 + (2d-3)\l_2}{d-1} \tr(K'^2) + 4 \l_1 \frac{d-3}{d-1} \nabla^n K^{ij} (\nabla_n K_{ij} - \nabla_j K_{in}) \nn \\
&+\Big(-4\l_1-\frac{2 \l_2}{d-1} \Big) \Big( \nabla_j K^{ij} (\nabla^n K_{in} - 2 \nabla_i \tr(K)) + \nabla_i \tr(K) \nabla^i \tr(K) \Big) \nn \\
& + \Big( \frac{2 d \l_2 + (22-10d) \l_3}{d-1} + 16 \l_1 \Big) \tr(K') \tr(K^2) + \Big(\frac{2(d-2) \l_2}{d-1} + 4\l_1 \Big) \tr(K) \tr(K K') \nn \\
&+\Big(-12 \l_1 -\frac{4(d-2) \l_2}{d-1} \Big) \tr(K) \tr(K^3) + \frac{(8-2d) \l_3 - \l_2}{d-1} \tr(K)^2 \tr(K^2)  \nn \\
& + \frac{3(2d-5) \l_3 -(2d-1)\l_2 - 2(6d-5)\l_1}{d-1} \tr(K^2)^2 +\frac{2(d-3)\l_3}{d-1} \tr(K') \tr(K)^2 \nn \\
& -\frac{2\l_1 + \l_2 + 2 \l_3}{d-1} \Big(d \pa_r^2 + \tr(K) \pa_r + \Box_{\g} \Big) \Big( \tr(3K^2 - 2K') - \tr(K)^2 +R \Big) \nn \\
& + \frac{4(9-7d) \l_1 + 4(3-2d)\l_2}{d-1} \tr(K^2 K') + 2 \l_3 R \tr(K^2-K') \nn \\
& + (4 \l_1 + \l_2) \Big( (\pa_r^2 + \tr(K) \pa_r + \Box_{\g}) \tr(K^2 -K') - (2 \nabla_i K^{ij} + 4 K^{ij} \nabla_i)(\nabla^n K_{jn} - \nabla_j \tr(K)) \Big) \Big]. \nn
\end{align}

Taking the trace of \eqref{fieldeqs} we obtain
\begin{align}
&0=R + \tr(3 K^2 - 2 K') - \tr(K)^2 + \frac{d(d+1)}{L^2} + \\
+& \frac{d-3}{d-1} L^2 \Big[\l_1 \Big(Riem_{\g}^2 + 4 R^{ijmn}K_{jm} K_{in} + 8 \nabla^n K^{ij} (\nabla_n K_{ij} - \nabla_j K_{in}) \Big) \nn \\
& + \l_2 \Big( Ric_{\g}^2 + 2(\nabla^i \tr(K)-\nabla_n K^{in})(\nabla_i \tr(K) - \nabla^j K_{ij}) + 2 R^{ij}(2(K^2)_{ij} - K'_{ij} - \tr(K)K_{ij}) \Big) \nn \\
& +\l_3 \Big(R_{\g}^2 + 2 R_{\g} (\tr(3K^2 - 2 K') - \tr(K)^2) \Big) + (2\l_1 + \l_2 + 9\l_3)\tr(K^2)^2 +2(\l_1 + 2 \l_2)\tr(K^4) \nn \\
& + (4\l_1 + \l_2) \tr(K'^2) - 4(2\l_1 + \l_2) \tr(K' K^2) + (\l_2 + 4 \l_3) \tr(K')^2 - 2 (\l_2 + 6 \l_3) \tr(K') \tr(K^2) \nn \\
&+ 2 \l_2 \tr(K) \tr(KK' - 2 K^3) + (\l_2 - 6 \l_3) \tr(K^2) \tr(K)^2 + 4 \l_3 \tr(K') \tr(K)^2 + \l_3 \tr(K)^4 \nn \\
& -\frac{1}{d-3}\Big(4 \l_1 + (d+1) \l_2 + 4 d \l_3\Big) \Big(\pa_r^2 + \tr(K) \pa_r + \Box_{\g}\Big) \Big(R + \tr(3 K^2 - 2 K') - \tr(K)^2 \Big) \Big]. \nn
\end{align}

The $(ij)$ components of the Einstein equations \eqref{fieldeqs} is
\begin{align}
&0 = R_{ij} - K'_{ij} + 2(K^2)_{ij} - \tr(K) K_{ij} + \frac{d}{L^2} \g_{ij} + \\
& + L^2 \Big[ \frac{1}{2(d-1)} \r^2 \g_{ij} +2(4\l_1 + \l_2) \tr(K^2) (K^2)_{ij} + (2\l_1 + \l_2) \tr(K'-K^2) K'_{ij} \nn \\
& + 2 \l_1 \Big(\frac{1}{2} R_{ilmn}R_j{}^{lmn} + 2 K^{mn} K_j{}^l R_{imnl} - (R^2)_{ij} + \nabla_i K^{mn} \nabla_j K_{mn} -\nabla^m K_{im} \nabla^n K_{jn} \nn \\
&+ \nabla^n K_i{}^m(2 \nabla_n K_{jm} - \nabla_m K_{jn} - 2\nabla_j K_{mn}) - \nabla_i \tr(K) \nabla_j \tr(K) + 2 \nabla_i \tr(K) \nabla^n K_{jn} \Big) \nn \\
& - 2(3\l_1 + \l_2) \tr(K') (K^2)_{ij} - 4(3\l_1 + \l_2) (K^2 K')_{ij} - \l_2 (RK')_{ij} + 2 \l_2 (K^2 R)_{ij} \nn \\
& + 2(2 \l_1 + \l_2) \tr(K) (K K')_{ij} + (2 \l_1 + \l_2) \tr(K)^2 (K^2)_{ij} - 2(7 \l_1 + 3 \l_2) \tr(K) (K^3)_{ij} \nn \\
& +(4 \l_1 + \l_2) \Big(\frac{1}{2}(\pa_r^2 + \tr(K) \pa_r + \Box_{\g}) (R_{ij} - K'_{ij} + 2(K^2)_{ij} - \tr(K) K_{ij} ) + \g^{mn} K'_{im} K'_{jn} \nn \\
& - 2 K_i{}^n \pa_r (R_{jn} - K'_{jn} + 2(K^2)_{jn} - \tr(K) K_{jn}) + (\nabla^n K_{jn} + 2 K_j{}^n \nabla_n)(\nabla^m K_{im} - \nabla_i \tr(K))\Big) \nn \\
& + 2 (3\l_1 + \l_2) K_i{}^n(R_{mn} - K'_{mn} + 2(K^2)_{mn}) K_j{}^m  + 4(2 \l_1 + \l_2) (K^4)_{ij} - \l_2 \tr(K) (KR)_{ij} \nn \\
& -\frac{1}{2}(2\l_1 + \l_2 + 2\l_3) (K_{ij} \pa_r + \nabla_i \nabla_j)\Big(R + \tr(3K^2 - 2 K') -\tr(K)^2 \Big) \nn \\
&-(2\l_1 + \l_2) \Big(R^m{}_{ij}{}^n (R_{mn} - K'_{mn} + 2(K^2)_{mn} - \tr(K) K_{mn}) +\tr(2 K^3 - KK' + KR) K_{ij} \nn \\
& \qquad \qquad \quad - 2(\nabla^n K_{ij} - \nabla_{(i} K_{j)}{}^n) (\nabla_n \tr(K) - \nabla^m K_{mn}) - \tr(K) \tr(K^2) K_{ij} \Big)  \nn \\
& + \l_3 (R + \tr(3K^2 - 2 K') -\tr(K)^2  ) \Big(R_{ij} - K'_{ij} + 2(K^2)_{ij} - \tr(K) K_{ij} \Big)  + (i \leftrightarrow j)\Big], \nn
\end{align}
where
\begin{align}
\r^2 =& -\l_1 Riem^2[G] - \l_2 Ric^2[G] - \l_3 R^2[G] -(2 \l_1 + \l_2 + 2\l_3) \Box_G R[G] \\
 =& -(\l_1 Riem_{\g}^2 + \l_2 Ric_{\g}^2) + 2 \l_2 R^{ij} (K'_{ij} - 2(K^2)_{ij} + \tr(K) K_{ij}) - 4 \l_1 R^{ijmn} K_{jm} K_{in} \nn \\
& - \l_3 \Big(R^2 + 2 R(\tr(3 K^2 - 2 K') - \tr(K)^2) \Big) -8 \l_1 \nabla^n K^{ij} (\nabla_n K_{ij} - \nabla_j K_{in}) -\l_3 \tr(K)^4 \nn \\
&- 2 (\l_1 + 2 \l_2) \tr(K^4) - (4 \l_1 + \l_2) \tr(K'^2) + 4(2 \l_1 + \l_2) \tr(K' K^2) - (\l_2 + 4 \l_3)\tr(K')^2  \nn \\
& - 2\l_2 \Big(\nabla_n K^{in} (\nabla^j K_{ij} - 2 \nabla_i \tr(K)) + \nabla_i \tr(K) \nabla^i \tr(K) \Big) + 2 (\l_2 + 6 \l_3) \tr(K') \tr(K^2)\nn \\
& -(2 \l_1 + \l_2 + 9\l_3) \tr(K^2)^2 - 2 \l_2 \tr(K) \tr(K K' - 2 K^3) + (6\l_3 - \l_2) \tr(K)^2 \tr(K^2) \nn \\
&- 4 \l_3 \tr(K') \tr(K)^2 \nn \\
& - (2 \l_1 + \l_2 + 2\l_3)(\pa_r^2 + \tr(K) \pa_r + \Box_{\g}) \Big( R + \tr(3K^2 - 2 K') -\tr(K)^2  \Big). \nn
\end{align}

\bibliographystyle{JHEP} 
\bibliography{literature}

\end{document}